# On A New Formulation of Micro-phenomena: The Double-slit Experiment

A. Shafiee[1] (*), A. Massoudi[2] (**)
and M. Bahrami[3] (*)

*) Research Group On Foundations of Quantum Theory and Information,
Department of Chemistry, Sharif University of Technology,
P.O.Box 11365-9516, Tehran, Iran.

**) Center of Complex Systems Research, Department of Chemistry,
K N Toosi University of Technology,
P. O. Box 1587-4416, Tehran, Iran

**Abstarct**

Based on the novel view that a micro-entity could be considered as a particle associated with a field partaking of the energy of particle which are both described by deterministic causal equations of motion, we examine the success of our new theory in elucidating the underlying physics of the double-slit experiment. Here, we explain with clear details how each micro-*particle* scatters from one of the slits at a given time. After the scattering through one of the slits, the particle shares some of its energy with its surrounding field and a particle-field system is again formed which its motion is governed by a deterministic dynamics during its flight towards the detecting screen. The interference pattern is then explained by showing how the final location of each particle-field system at the time of reaching the detecting screen is distributed according to an angular distribution (equal to the what quantum theory predicts for the fringe effects in a two-slit experiment). The probabilistic nature of such a distribution can be explained by considering the variations of the kinetic energy of the particle-field system at different local situations.

## 1   Introduction

The wave-like nature of light was first proved by Thomas Young who observed the interference pattern of light in his famous double-slit experiment in1802 [1]. However, if we think of light as consisting of photons, the interference pattern of light will seem mysterious. Nevertheless, this is not a characteristic feature of light, only. After Young's examination, we know at present that the double-slit experiment demonstrates the strange wave behavior of micro-entities, not merely the photons. Yet, there has been provided no physical theory thus far which could explain the interference phenomenon in a two-slit experiment on the basis of a *single* micro-entity foundation. For this reason, Feynman states that "[i]n reality, it contains the *only* mystery" in the quantum domain [2]. The main mysterious point is that in the two-slit experiments, it appears as though each single entity interferes with itself. Because, if only one micro-entity (say, an electron) is sent at a time from the source towards the slits, the interference fringes are also observed when a sufficiently large number of micro-entities has been counted. But, according to a particle picture, e.g., an electron which has been sent from a source, either passes through one of the slits or passes through another slit. If this were the correct description, we would be unable to account for the interference fringes. For, if the electron passes through one of the slits, the other slit can be neglected and it should be possible for the electron to arrive at all points on the detecting screen after the slits. No bright and dark fringes will happen afterwards. If one still challenges this view by detecting through which slit the electron actually passes, the interference fringes will be destroyed. Any possibility which enables us to know through which slit the micro-entity has been passed perturbs the whole quantum state of the system to such an extent that no interference pattern is detected. In effect, two complementary events can be recognized here: *either* we know which slit the micro-entity passed through, *or* we observe the interference effect.

---

[1] Corresponding author: shafiee@sharif.edu
[2] a_massoudi@alborz.kntu.ac.ir
[3] mbahrami@mehr.sharif.ir

In this paper, we are going to give an explanation for the puzzling problem of interference fringes by using the causal formulation of micro-incidents described in [3] and [4]. According to our approach, each micro-entity constitutes a particle which is associated with a field. A particle-field (PF) is an integrated system which can be imagined as a particle which its properties is merged with its surrounding space so much so it cannot be realized as a classical particle. What describes the surrounding space of a micro-particle is a probability field partaking of the energy of the particle. Any micro-particle shares its properties with its enfolding probability field to create a new meaning of a tiny substance. So, contrary to the de Broglie-Bohm depiction of micro-particles and pilot waves [5], in our description there is no room for *active* waves which can act on the particles (through a so-called quantum force) or guide their motion. A PF system should not be envisaged as a particle guided by a field. Rather, it should be considered as a unified entity with its own dynamics which is, however, affected by the physical conditions the *particle* is subjected to [3]. Hence, the particle plays the key role here and the behavior of its associated field is contingent upon what the particle sees and experiences. Nevertheless, the whole behavior of the entire system is very different from what we may observe for an individual particle alone, due to the interplay of the particle and its allied field.

In this way, we are going to elucidate that despite a PF system reaches one of the slits at a given time according to a determinate trajectory, it will arrive at specific locations on the detecting screen, causing the dark and bright interference fringes.

In the following section we will describe the physical situation of the experiment and explain the problem in details. Then, we have a concluding section to sum up our results.

## 2 The double-slit experiment

We prepare a source emitting a beam of micro-entities (electrons, neutrons, atoms or in general PF systems) towards a plane including two slits each of finite width $a$ separated by a distance $d$. The slits are placed along the $y$-direction and we ignore the thickness of them along the $x$ and $z$ directions. So, the $x$ and $z$ coordinates of the position of the slits are fixed in the $xz$ plane, denoted by $x_0$ and $z_0$, respectively. The intensity of the incident beam is supposed to be so low that the PF entities arrive at the slits one at a time. Every PF system sent from the source has a definite amount of linear momentum $p^0$ at $t < 0$ before reaching the slits, where (see Appendix A):

$$p^{0^2} = p_P^{0^2} A_0^2 > p_P^{0^2} \tag{1}$$

Here, $A_0$ is a scaler function of the components of the de Broglie momentum $p^0$ and the particle's momentum $p_P^0$, both defined at $t < 0$, as well as the amplitude of the momentum (plane) field $A_p$. The plane field is defined as:

$$\chi_p(x(t), y(t), z(t)) = A_p \exp(i\vec{k}^0 \cdot \vec{r}(t)); \quad t < 0 \tag{2}$$

In relation (2), $\vec{r}(t)$ is the local vector of a particle of mass $m$ in the Euclidean space and $k^0 = \frac{p^0}{\hbar}$, where $p^0$ is the magnitude of the de Broglie momentum defined in (1). The PF systems have no interaction with their environment, so the energy of each PF system is conserved and equal to its kinetic energy defined as $\frac{p^{0^2}}{2m}$ at $t < 0$.

If we assume that the momentum of each PF system has the same magnitude in all directions at $t < 0$ (that is $p^0$ is an *isotropic* momentum), using the relation $p_\beta^0 = p_{P,\beta}^0 A_0$ where $\beta = x, y, z$ and $A_0$ is a scaler function, we can conclude that

$$p_{P,\beta}^0 = \frac{\pm 1}{\sqrt{3}}\left(\frac{p^0}{A_0}\right) = \frac{\pm 1}{\sqrt{3}A_0}\left(\frac{h}{\lambda_0}\right) \tag{3}$$

where $\lambda_0 = \frac{h}{p_0}$ is the de Broglie wavelength interpreted in our theory as the associated wavelength of particle at $t < 0$ (see also appendix A).

Now, we assume that each PF entity, encompassing a particle enfolded by its associated plane field in (2), reaches the double-slit apparatus at $t = 0$, where the slits are placed along the $y$-direction. A system of slits is a position-measuring device [6, 7]. Thus, as stated earlier in [4], when a PF system reaches a slit, a discontinuity happens in the trajectory of the system and the dynamics suddenly breaks at $t = 0$. Hence, at the position of the slits, the energy of field is transferred to the particle under an irreversible conversion process resulting to an unfolding particle having a null field with zero amplitude. The function-dependency of the field, however, specifies the corresponding wave function of the particle at $t = 0$.

Accordingly, at the position of the slits, if we introduce the $y$-component of the particle's momentum as $p_{P,y}^S$, we have:

$$p_{P,y}^S = p_y^0 = \frac{\pm 1}{\sqrt{3}}\left(\frac{h}{\lambda_0}\right) \tag{4}$$

where $p_y^0$ is the de Broglie momentum of the PF system at $t < 0$. In reaching the relation (4), it is important to note that at $t = 0$, $A_p \to 0$ and $A_0 \to 1$ (see the Appendix A for the definition of $A_0$).

At $t = 0$ (i.e., when a PF entity arrives at the slits), the uncertainty relation of position and momentum only reflects our lack of knowledge about the exact values of $p_{P,y}^S$ and the $y$-component of the position of particle, since the particle's properties are well-defined at the hidden level. Assuming that a particle passing through the first slit, e.g., has a definite position $y$ in the continuum $y_1' - a/2 \le y \le y_1' + a/2$, the average values of $\langle y \rangle$ and $\langle y^2 \rangle$ can be obtained, respectively, as

$$\langle y \rangle = \frac{1}{a}\int_{y_1'-a/2}^{y_1'+a/2} dy\; y = y_1' \tag{5}$$

and

$$\langle y^2 \rangle = \frac{1}{a} \int_{y_1'-a/2}^{y_1'+a/2} dy \, y^2 = y_1'^2 + \frac{a^2}{12} \tag{6}$$

where $y_1'$ denotes the middle point of the first slit. We have also supposed that the particle has a uniform probability of being found through the slit. From the relations (5) and (6), one concludes that $\Delta y = \frac{a}{2\sqrt{3}}$. At the same time, according to (4), we have $\langle p_{P,y}^s \rangle = 0$ and $\langle p_{P,y}^{s^2} \rangle = \frac{h^2}{3\lambda_0^2}$ which shows that $\Delta p_{P,y}^s = \frac{h}{\sqrt{3}\lambda_0}$.[4] Therefore, from the indeterminacy relation $\Delta y \Delta p_{P,y}^s \geq \hbar/2$, one infers that

$$a \geq \frac{3}{2\pi} \lambda_0 \tag{7}$$

which states that the width of the slit cannot be smaller than about the half magnitude of the de Broglie momentum of the particle, in accordance with all theoretical as well as experimental considerations reported so far (see below).

At the position of the slits, the wave function of the *particle* can be defined as

$$\psi(y(t)) = \frac{1}{\sqrt{2}} \sum_{i=1}^{2} \psi_i(y(t)); \quad 0 \leq t \leq \varepsilon \tag{8}$$

where $\varepsilon$ is small enough to assume that a little after the slits, the spatial distribution of the particle along the $y$-direction can still be nearly given by $|\psi(y)|^2$. This assumption is essential, because for obtaining the field representation of the system at latter times in our approach, we need the trajectory equation of the particle after the slits which can be traced back to an initial time $t = \varepsilon \neq 0$, due to the discontinuity occurred at the position of the slits at $t = 0$.

In (8), we have

$$\psi_i(y(t)) = \begin{cases} \frac{1}{\sqrt{a}} & y_i' - a/2 \leq y \leq y_i' + a/2 \\ 0 & \text{elsewhere} \end{cases}; \quad 0 \leq t \leq \varepsilon \tag{9}$$

where $y_i'$ is the middle point of the $i$th slit ($i = 1$ or $2$). According to what argued before in [4], the amplitude of the field becomes zero, when the particle passes through the slit. So, the associated field of the particle at $t = 0$ is zero, demonstrating that our lack of knowledge about the physical situation of the particle is only a subjective matter with no objective feature [4]. After the passage of the particle through the slit at $t > 0$ (including $t \geq \varepsilon$), however, the field will partake of some of energy of the particle and its amplitude will be not zero.

---

[4]Each slit can be imagined as an one-dimensional box along the $y$-direction. Hence, the mean value of $\langle p_{P,y}^{s^2} \rangle$ is, strictly speaking, equal to $\frac{n^2 h^2}{4a^2}$ at $t = 0$, where $n$ is the energy quantum number. Assuming an uniform probability distribution for each slit along the $y$-direction, it is legitimate to consider that the magnitude of $n$ is so large that the value of $\langle p_{P,y}^{s^2} \rangle$ can be given by $\frac{h^2}{3\lambda_0^2}$ as a continuous quantity. For example, for the values of $\lambda_0 \approx 2nm$ and $a = 22\mu m$ in a double-slit neutron experiment [8], the value of $n$ can be approximated as $2 \times 10^{16}$.

At $t \geq \varepsilon$, the time evolution of the wave function in the $y$-direction can be described in terms of a free-particle propagator $K_y(y(t), t; y(\varepsilon), \varepsilon)$ defined as [4]:

$$K_y(y(t), t; y(\varepsilon), \varepsilon) = \sqrt{\frac{m}{i\hbar(t-\varepsilon)}} \exp\left[\frac{im(y(t)-y(\varepsilon))^2}{2\hbar(t-\varepsilon)}\right] \tag{10}$$

where $y(\varepsilon)$ is the $y$-component of the position of the particle at $t = \varepsilon$. The wave function of the PF system at $t \geq \varepsilon$ can now be defined as:

$$\Psi_y(y(t), t) = \frac{1}{\sqrt{2}} \sum_{i=1}^{2} \int_{y_i'-a/2}^{y_i'+a/2} dy(\varepsilon) \; K_y(y(t), t; y(\varepsilon), \varepsilon) \psi_i(y(\varepsilon)); \quad t \geq \varepsilon \tag{11}$$

At $t \to \varepsilon$, the propagator in (10) tends to $\delta(y - y(\varepsilon))$ and the wave function in (8) is reproduced for $t = \varepsilon$. As stated above, however, the associated field of the particle at $t > 0$ is endowed with energy. Thus, the wave function in (8) is indeed an approximation for $0 < t \leq \varepsilon$.

After passing through a given slit, the equation of motion of the particle along the $y$-direction can be expressed as:

$$y(t) - y(\varepsilon) = (t - \varepsilon)\frac{p_{P,y}}{m}; \quad t \geq \varepsilon \tag{12}$$

where $p_{P,y}$ is the $y$ component of the momentum of the particle for $t \geq \varepsilon$. Regardless of what slit the particle has passed through, we also assume that the particle's momentum $p_{P,y}$ is the same value.

Using the relations (10)-(12), the relation (11) can now be given as follows:

$$\Psi_y(y(t), t) = \frac{1}{\sqrt{2a}} \sum_{i=1}^{2} \sqrt{\frac{m}{i\hbar(t-\varepsilon)}} \int_{y_i'-a/2}^{y_i'+a/2} dy(\varepsilon) \; \exp\left[\frac{ip_{P,y}(y(t)-y(\varepsilon))}{2\hbar}\right] \tag{13}$$

Then, $\Psi_y(y(t), t)$ in (13) can be represented as:

$$\Psi_y(y(t), t) \propto \left(\frac{\sin\alpha}{\alpha}\right) \frac{\exp\left(\frac{ip_{P,y}y(t)}{2\hbar}\right)}{\sqrt{t-\varepsilon}} \sum_{i=1}^{2} \exp\left(-\frac{ip_{P,y}y_i'}{2\hbar}\right); \quad t \geq \varepsilon \tag{14}$$

where

$$\alpha = \frac{ap_{P,y}}{4\hbar} \tag{15}$$

Now, for $t \geq \varepsilon$, one can introduce the field representation of the PF system in the following form[5]:

---

[5] Here, while we have a continuous change of state from $\psi(y(0))$ to $\Psi(y(t), t)$ (including $t = \varepsilon$), the corresponding field of the system changes discontinuously from a null field at $t = 0$ to $X(y(t), t) \neq 0$ for $t \geq \varepsilon$. For this reason, we have traced the time evolution of the wave function according to the time-dependent dynamics of the system, but assign the field representations at two different time stages $t = 0$ and $t \geq \varepsilon$, inferentially.

$$X_y(y(t),t) = \left(\frac{\sin\alpha}{\alpha}\right) F_y(t) \sum_{i=1}^{2} \exp\left(-\frac{ip_{P,y}y'_i}{2\hbar}\right); \quad t \geq \varepsilon \tag{16}$$

where

$$F_y(t) = \frac{A_y(t)}{\sqrt{(t-\varepsilon)}} \exp\left(\frac{ip_{P,y}y(t)}{2\hbar}\right) \tag{17}$$

and $A_y(t)$ is a time-dependent factor. Since it is supposed that each particle and its associated field are free of any interaction during the flight from the source towards the detecting screen, one can assume that:

$$\frac{d|\dot{X}_y|}{dt} \propto \frac{d|\dot{F}_y(t)|}{dt} = 0 \tag{18}$$

Then, it can be resulted from (18) that $A_y(t) = c_{F,y}\sqrt{t - \varepsilon}$, where $c_{F,y}$ is a real constant [4]. Thus, one can finally obtain the field representation of the PF system as $X(y(t), t) \rightarrow \chi(y(t))$, where

$$\chi_y(y(t)) = c_{F,y}\left(\frac{\sin\alpha}{\alpha}\right) \exp\left(\frac{ip_{P,y}y(t)}{2\hbar}\right) \sum_{i=1}^{2} \exp\left(-\frac{ip_{P,y}y'_i}{2\hbar}\right); \quad t \geq \varepsilon \tag{19}$$

The situation is, however, different along the $x$ and $z$ directions. When the particle passes through one of the slits located along the $y$-direction, the $x$ and $z$ components of its position will take a definite value (denoted as $x_0$ and $z_0$, respectively), if we neglect the thickness of the slits along these directions. Thus the wave function of the particle in the $x(z)$-direction can be represented as $\delta(x(\varepsilon) - x_0)$ ($\delta(z(\varepsilon) - z_0)$) at $0 \leq t \leq \varepsilon$ for small $\varepsilon$. The associated field of the particle is null at this time interval in both directions $x$ and $z$.

Now, let us focus on the $x$-direction in the meantime. The wave function of the system in the $x$-direction at the position of the slits can be given as:

$$\Psi_x(x(t),t) = \int_{-\infty}^{+\infty} dx(\varepsilon) \, K_x(x(t),t;x(\varepsilon),\varepsilon)\delta(x(\varepsilon) - x_0); \quad t \geq \varepsilon \tag{20}$$

where $K_x(x(t), t; x(\varepsilon), \varepsilon)$ is the free particle propagator along the $x$-direction. Considering the motion of the particle along the $x$-direction as:

$$x(t) = (t - \varepsilon)\frac{p_{P,x}}{m} + x(\varepsilon); \quad t \geq \varepsilon \tag{21}$$

one can express the relation (20) in the form of:

$$\Psi_x(x(t),t) = \sqrt{\frac{m}{i\hbar(t-\varepsilon)}} \exp\left[\frac{ip_{P,x}(x(t)-x_0)}{2\hbar}\right]; \quad t \geq \varepsilon \tag{22}$$

where $p_{P,x}$ is the $x$ component of the linear momentum of the particle at $t \geq \varepsilon$. After passing through one of the slits, the particle shares energy with its surrounding and the field representation becomes significant. Thus, using the no-physical-interaction hypothesis for the PF system departing the slits, one can assign the field representation in the $x$-direction as:

$$\chi_x(x(t)) = c_{F,x}\exp\left[\frac{ip_{P,x}(x(t)-x_0)}{2\hbar}\right]; \quad t \geq \varepsilon \tag{23}$$

where $c_{F,x}$ is a real constant. A similar relation can be obtained for particle's field in $z$-direction. The entire field of the PF system, however, can be factorized:

$$\chi(x,y,z) = \chi_x(x(t))\chi_y(y(t))\chi_z(z(t)) \tag{24}$$

where $\chi_x(x(t))$ and $\chi_y(y(t))$ are given in (23) and (19), respectively, and $\chi_z(z(t))$ has the same form as $\chi_x(x(t))$.

According to (24), the probability field can now be presented as the follwoing:

$$|\chi(\theta)|^2 = 4c_F^2 \left(\frac{\sin\alpha}{\alpha}\right)^2 \cos^2\left(\frac{\phi}{2}\right) \tag{25}$$

where $c_F = c_{F,x}c_{F,y}c_{F,z}$ is the amplitude of the entire field with dimension of length. In (25), $\phi = \frac{p_{P,y}d}{2\hbar}$ and $d = |y_2' - y_1'|$ is the distance between the slits, center to center. We also define $p_{P,y} = p_P\sin\theta$, where $\frac{\pi}{2} \leq \theta \leq -\frac{\pi}{2}$ is the scattering angle of the particle which for simplicity is assumed to have a specific value, regardless of what slit the particle has been scattered through. Correspondingly, the angular probability distribution for the scattering of the particle can be represented as:

$$|\psi(\theta)|^2 = \mathcal{N}^2 \left(\frac{\sin\alpha}{\alpha}\right)^2 \cos^2\left(\frac{\phi}{2}\right) \tag{26}$$

where $\mathcal{N}$ is a normalization constant. The probability distribution (26) is similar to what quantum mechanics predicts for the probability that a particle is being scattered from a two-slit apparatus [7]. The only difference is that the scattering angle $\theta$ here is twice the scattering angle defined in the quantum mechanics (see below). Hence, $\theta = 2\theta_{QM}$ and $\sin\theta \approx 2\sin\theta_{QM}$, due to the smallness of the scattering angle.[6] It is also interesting to note that, if we have only one slit (assuming $\phi = 0$ in (20), since $d = 0$), the diffraction form $\left(\frac{\sin\alpha}{\alpha}\right)^2$ will be obtained which is the well-known diffraction distribution for photons and other quantum particles [7].

The kinetic energy of the field, for $t \geq \varepsilon$, is equal to:

$$K_F = \frac{1}{2}m|\dot{\chi}|^2 = \frac{1}{2m\hbar^2}p_P^4 c_F^2 \left(\frac{\sin\alpha}{\alpha}\right)^2 \cos^2\left(\frac{\phi}{2}\right) \tag{27}$$

---

[6]The reader should note that the scattering angle in our approach has a clear meaning, because we present a real picture of the scattering process here. In quantum mechanics, however, the definition of $\theta$ is vague, since we do not know how micro-entities can be scattered by the system of slits to produce the interference pattern collectively.

where $p_P = \left( \sum_{\beta=x,y,z} p_{P,\beta}^2 \right)^{1/2}$ is the magnitude of the momentum of particle at $t \geq \varepsilon$. The momentum of the PF system can now be obtained as:

$$p = m\dot{q} = (p_P^2 + 2mK_F)^{1/2} = p_P \left[ 1 + \frac{p_P^2}{4\hbar^2} |\chi(\theta)|^2 \right]^{1/2} \tag{28}$$

where $|\chi(\theta)|^2$ is defined in (25).

Due to the conservation of the energy, we can relate the kinetic energy of the PF system before and after the slits, according to the following relation

$$p^2 = \sum_{\beta=x,y,z} p_\beta^2 = p^{0^2} = \frac{h^2}{\lambda_0^2} \tag{29}$$

where $p^0$ is defined in (1) and $\lambda_0$ is the associated wavelength of the particle at $t < 0$ (see also relation (3)). Using the relation (28), one gets

$$p_P^2 = \frac{-1 + \sqrt{1 + \frac{p^{0^2}}{\hbar^2}|\chi(\theta)|^2}}{\frac{1}{2\hbar^2}|\chi(\theta)|^2} \tag{30}$$

According to (28), we can define:

$$p_\beta = p_{P,\beta} \left[ 1 + \frac{p_P^2}{4\hbar^2} |\chi(\theta)|^2 \right]^{1/2} \tag{31}$$

where $p_{P,x} = p_P \sin\vartheta \cos\theta$, $p_{P,y} = p_P \sin\vartheta \sin\theta$, $p_{P,z} = p_P \cos\vartheta$ and $p_P$ is defined in (30). Henceforth, we take $\vartheta \approx \frac{\pi}{2}$, considering only two components of the PF's position in the $xy$ plane for simplicity. By using the relation (30), it is straightforward to show that in (31):

$$p_x = p_P^0 A_0 \cos\theta = \frac{h}{\lambda_0} \cos\theta \tag{32}$$

and

$$p_y = p_P^0 A_0 \sin\theta = \frac{h}{\lambda_0} \sin\theta \tag{33}$$

where $-\frac{\pi}{2} \leq \theta \leq \frac{\pi}{2}$. From the relations (32) and (33), at $t \geq \varepsilon$ we can obtain the $x$ and $y$ components of $q$, respectively, as:

$$q_x = A_0 v_P^0 (t - \varepsilon)\cos\theta + x_0; \quad t \geq \varepsilon \tag{34}$$

and

$$q_y = A_0 v_P^0(t-\varepsilon)\sin\theta + y(\varepsilon); \quad t \geq \varepsilon \tag{35}$$

where $y(\varepsilon)$ is the $y$ component of the position at $t = \varepsilon$. All the equations (32)-(35) are obtained in an exact manner. Nevertheless, assuming that the energy of field is not so great, so that only the second power of the field's amplitude is significant, one can show that in (30) $p_P^2; p^{0^2} = \frac{h^2}{\lambda_0^2}$.[7] Thus, we can approximately express the parameters $\alpha$ and $\phi$ in (25) and (26) as:

$$\alpha = \frac{a p_{P,y}}{4\hbar}; \frac{a\pi\sin\theta}{2\lambda_0} \tag{36-a}$$

and

$$\phi = \frac{d p_{P,y}}{2\hbar}; \frac{d\pi\sin\theta}{\lambda_0} \tag{36-b}$$

which are similar to the quantum definitions, if we consider $\sin\theta = 2\sin\theta_{QM}$ [7].

To see how we can interpret the relation (26) as a density function describing the probability of observing the positions of the particles along the $y$-direction, we examine the kinetic energies of the system at different situations [3]. First, one should note that when $\sin\theta = 0$, $K_F$ in (27) is positive and maximum, but the $y$ component of the particle's momentum is zero. Correspondingly, $K_{PF,y} = \frac{p_y^2}{2m} = 0$, showing that the it is more likely to find the *particle* in such region and the probability of finding the particle in the $y$-direction will be maximum. This is also in accordance with what the probability distribution $|\psi(\theta)|^2$ describes in (26), when $\theta = 0$.

On the other hand, for regions where $\sin\theta \neq 0$, but $|\chi(\theta)|^2 = 0$, $K_F$ in (27) becomes zero and the momentum of the particle in (30) reaches its maximum value of $\frac{h}{\lambda_0}$. Hence, $p_{P,y}$ approaches its maximum value of $p_y$ in (33). So, the chance of observing the particle in such regions vanishes.

For regions where $\sin\theta \neq 0$ and also $|\chi(\theta)|^2 \neq 0$, $K_F$ has a limited nonzero value in (27), demonstrating that the kinetic energy of the particle has been diminished in such regions, but still is significant. Correspondingly, the chance of finding the particle in such regions is probable to some extent.

Eventually, at $t = T$, we assume that the PF system bumps into a detecting screen which is located at distance $L$ from the slits. Subsequently, a position measurement takes place at $t = T$ and a *particle* is detected on the screen. For, when the position of a micro-entity is measured, the position of the PF system is reduced to particle's siting, according to an irreversible discontinuous process [4]. Thus, e.g., we can never measure the $q_y$ introduced in (35) (or other spatial coordinates of the PF system) at $T$. In other words, PF's spatial coordinates are not *discernible* in practice. Rather, every spot on the detecting screen signifies the location of a detected particle denoted by some $x_{\text{det}}(T)$, $y_{\text{det}}(T)$, and $z_{\text{det}}(T)$ values. Since this kind of transformation from the PF to the particle position cannot be followed up dynamically, the *exact* values of the particle's spatial coordinates on the detecting screen cannot be determined *a priori*. Nevertheless, the difference between each PF's spatial coordinates $q_x$, $q_y$ and $q_z$ with its corresponding components $x_{det}$, $y_{det}$ and $z_{det}$ for the particle at $t = T$ is not so significant, because the whole detected space (i.e., the space where the spots are observed) are so small that any discrimination between the PF and the particle spatial coordinates becomes negligible for practical purposes. As a matter of fact, the same pattern delineated by the spatial coordinates of a PF system can be then observed for the detected particles as well, but in a more compact region, because the PF system is a more extensive system comprised of both particle and its allied field. Consequently, having into account that the density function in (26) could be indeed interpreted as the probability distribution of the detected $y$ components of particles' positions (denoted by $y_{det}(T)$) and assuming that $q_y(T) \approx y_{det}(T)$, one can sketch the interference pattern of the particles scattered from the slits at $t = 0$.

---

[7]This is indeed a legitimate assumption taking notice that the energy of a PF system should be concentrated mainly on the particles itself after the slits, because the energy transfer from the particle to its surrounding is so fast that at $0 \leq t \leq \varepsilon$ the field cannot partake of a considerable energy.

Based on our results in this paper, in figures 1-a, b, c and 2-a, b, c, the cross section of the interference pattern for particles scattered from the two slits and its corresponding angular distribution $|\psi(\theta)|^2$ in (26) (corresponding to the probability field $|\chi(\theta)|^2$ in (25)), together with a graph of the impacts on the detecting screen are shown for two well-known two-slit experiments regarding electrons and ultracold neon atoms, respectively [9, 10]. For reproducing the purposed pattern depicted by $|\psi(\theta)|^2$, we should have $A_0 10$ (1) and $|\theta| \leq \frac{\pi}{50000}$ ($\leq \frac{\pi}{200}$) for the electron (neon atom) two-slit experiment which shows that for larger quantum particles the energy of the plane field before the slits can be neglected (i.e., in relations (34) and (35), $A_0 \to 1$). After the slits, however, the presence of field is important, because without an energetic field (with small energy, albeit), we have no reason why the PF coordinates in (34) and (35) should be distributed according to $|\psi(\theta)|^2$. We have also assumed that $v_P^0(T - \varepsilon) \approx L$. The length of the detecting screen (as is clear in figures 1-a, b) is about $0.16$ mm for the electron experiment. This length is about $3.2$ mm for the neon atom experiment (depicted in figures 2-a, b) which can be compared to the value of $\approx 2$ mm in real experiment, showing that the impacted particles have indeed a more compact pattern in practice than what PFs' trajectories describe in theory.

## 3   Conclusion

The total picture we have depicted here to explain the interference phenomenon constitutes of several substantial elements. First, we assumed that at $t < 0$ there is a source emitting free PF systems with low intensity, each PF system possesses a definite amount of linear momentum $p^0$. The field is described by a plane wave and partakes of some limited amount of particle's energy. At $t = 0$, a given PF system reaches the slits placed along the $y$-direction. Due to the conservation of energy, the continuous kinetic energy of the PF system remains unchanged, but its $y$ component specifies the high energy levels of a *particle* in one dimensional box in the $y$-direction (see the relations (3) and (4) for comparison). So, the energy of field becomes zero and the $y$ component of the position of particle inside the slit can be described by a uniform distribution $\frac{1}{a}$, where $a$ is the width of each slit. After passing through one of the slits, however, the associated field partakes of some limited amount of particle's energy again. This given energy has no direct effect on PF's trajectories at $t > 0$ (introduced in relations (34) and (35), e.g.), but determines the local distribution of PF's spatial coordinates, after the particle is scattered from a slit. If there was no energetic field after the slits, the spatial distribution of the particle would be specified by evaluating the variations of the kinetic energy of the particle in different locations. This, however, would lead to the conclusion that the particles should most frequently impact the detecting screen at positions located at the front of each slit (i.e., the places near the scattering angle $\theta = 0$). This is indeed the pattern observed for the scattered classical objects. Yet, for micro-entities the situation is different. Since the particle shares its energy with its surrounding, the kinetic energy of it varies at different locations depending upon the amount of energy its associated field is partaking of. This, in turn, explains why the spatial distribution of the detected spots at the final screen should be in accordance with what the angular distribution $|\psi(\theta)|^2$ describes. Consequently, one can also ascribe the same statistical weight to the trajectories of the PF system, because the locations these trajectories determine at the time of detection can be assumed to be nearly equal to the locations a *particle* might be detected concurrently.

In effect, the interference fringes can be interpreted as the fine structure of a classical pattern manifested for micro-entities. Neglecting the subtleties of an interference pattern, a Gaussian-type distribution of scattered particles can be roughly reconstructed. In the micro-domain, however, what causes the appearance of the fringes is the association of an energetic field in the whole process. This leads to a fine structure, because the amount of energy which the particle shares with its surrounding is small. For the macro-particles, however, the probability field has no contribution in the entire energy of the system, since as the particle becomes more discernible itself, the difference between its own attributes and the properties being observed in practice diminishes gradually. Accordingly, the interference phenomenon demonstrates the inherent feature of all micro-particles to share some of their energy with their surrounding to produce an allied field partaking of the same energy. In such a way, the mutual relationship between a micro-particle and its associated field leading to the appearance of an integrated new PF entity can coherently explain the mysterious interference fringes of the double-slit experiments.

**Appendix A**

For a plane field described as (2), the energy of field can be defined as:

$$E_F = K_F = \frac{1}{2}mv_F^{0^2} \tag{A-1}$$

and the velocity of field is given as

$$v_F^0 = \left|\vec{v}_P^0 \cdot \vec{\nabla}\chi_p\right| = \left|A_p \sum_{\beta=x,y,z} v_{P,\beta}^0 k_\beta^0\right| \tag{A-2}$$

where $v_P^0$ is the velocity of the particle at $t < 0$, $\vec{k}^0 = \frac{\vec{p}^0}{\hbar}$ and $\chi_p$ is defined in (2). Now, for the momentum of the PF system before reaching the slits, one can show that:

$$p^{0^2} = p_P^{0^2} + A_p^2\left(\sum_{\beta=x,y,z} p_{P,\beta}^0 k_\beta^0\right)^2 > p_P^{0^2} = p_P^{0^2} A_0^2 \tag{A-3}$$

where

$$A_0^2 = 1 + \left(\frac{A_p \sum_{\beta=x,y,z} p_{P,\beta}^0 k_\beta^0}{p_P^0}\right)^2 > 1 \tag{A-4}$$

According to (A-3), for each component of the de Broglie momentum $p^0$ as well as the particle's momentum $p_P^0$, one can define the relation $p_\beta^0 = p_{P,\beta}^0 A_0$ where $\beta = x, y, z$. For an isotropic de Broglie momentum, we have $k^{0^2} = 3k_\beta^2$ and the relation (A-3) can be written as:

$$p^{0^2} = \frac{p_P^{0^2}}{\left(1 - \frac{A_p^2 \pi_P^{0^2}}{3\hbar^2}\right)} \quad (A-5)$$

where

$$\pi_P^{0^2} = \left(\sum_{\beta = x,y,z} p_{P,\beta}^0\right)^2 \quad (A-6)$$

At $t < 0$, assuming an isotropic de Broglie momentum, we have $p_{P,\beta}^0 = \frac{\pm 1}{\sqrt{3}}\left(\frac{p^0}{A_0}\right)$. At $t = 0$, i.e., at the position of the slits along the $y$-direction, $A_p \to 0$ and $A_0 \to 1$, hence $p_{P,\beta}^s \to \frac{\pm 1}{\sqrt{3}} p^0 = p_\beta^0$.

## References

[1] T. Young, "On the theory of light and colors", *Philos. Trans. RSL*, **92**, 12-48 (1802).
[2] R. P. Feynman R. B. Leighton and M. Sands, *The Feynman Lectures on Physics*, (Addison-Wesely, Reading, 1965), Vol. 3.
[3] A. Shafiee, "On A New Formulation of Micro-phenomena: Basic Principles, Stationary Fields And Beyond", submitted article.
[4] A. Shafiee, " On A New Formulation of Micro-phenomena: Quantum Paradoxes And the Measurement Problem", submitted article.
[5] D. Bohm and B. J. Hiley, *The Undivided Universe*, (Routledge, London, 1993).
[6] C. Cohen-Tannoudji, B. Diu and F. Laloe, *Quantum Mechanics* (Wiley, New York, 1977), p. 261.
[7] T. V. Marcella, "Quantum interference with slits", *Eur. J. Phys.*, **23**, 615-621 (2002).
[8] A. Zeilinger, R. Gähler, C. G. Shull, W. Treimer, and W. Mampe, "Single and double slit diffraction of neutrons", *Rev. Mod. Phys.*, **60**, 1067-1073 (1998).
[9] C. Jönsson, "Elektroneninterferenzen an mehreren künstlich hergestellten feinspalten", *Z. Phys.*, **161**, 454-474 (1961). English translation: "Electron diffraction at multiple slits", *Am. J. Phys.*, **42**, 4-11 (1974).
[10] F. Shimizu, K. Shimizu, and H. Takuma, "Double-slit interference with ultracold metastable neon atoms", *Phys. Rev. A*, **46**, R17-R20 (1992).

## Figure Captions

Figure Caption 1 (a) The cross section of the interference pattern for a total number of 5000 electrons scattered from the two slits. The vertical axis shows the number of hits and the horizontal axis shows the length of the screen in terms of meter. (b) The corresponding angular distribution $|\psi(\theta)|^2$ with $|\theta| \leq \frac{\pi}{50000}$. The horizontal axis is in terms of radian. (c) The impacts on the detecting screen for the same number of electrons as in (a). Here, we have used the data $\lambda_0 = 5 \times 10^{-12}$ $m$, $d = 2$ $\mu m$, $a = 0.5$ $\mu m$, $L = 0.35$ $m$ [9] and $A_0 = 10$.

Figure Caption 2 (a) The cross section of the interference pattern for a total number of 5000 ultracold neons scattered from the two slits. The vertical axis shows the number of hits and the horizontal axis shows the length of the screen in terms of meter. (b) The corresponding angular distribution $|\psi(\theta)|^2$ with $|\theta| \leq \frac{\pi}{200}$. The horizontal axis is in terms of radian. (c) The impacts on the detecting screen for the same number of neons as in (a). Here, we have used the data $\lambda_0 = 1.8 \times 10^{-8}$ $m$, $d = 6$ $\mu m$, $a = 2$ $\mu m$, $L = 0.113$ $m$ [10] and $A_0 = 1$.

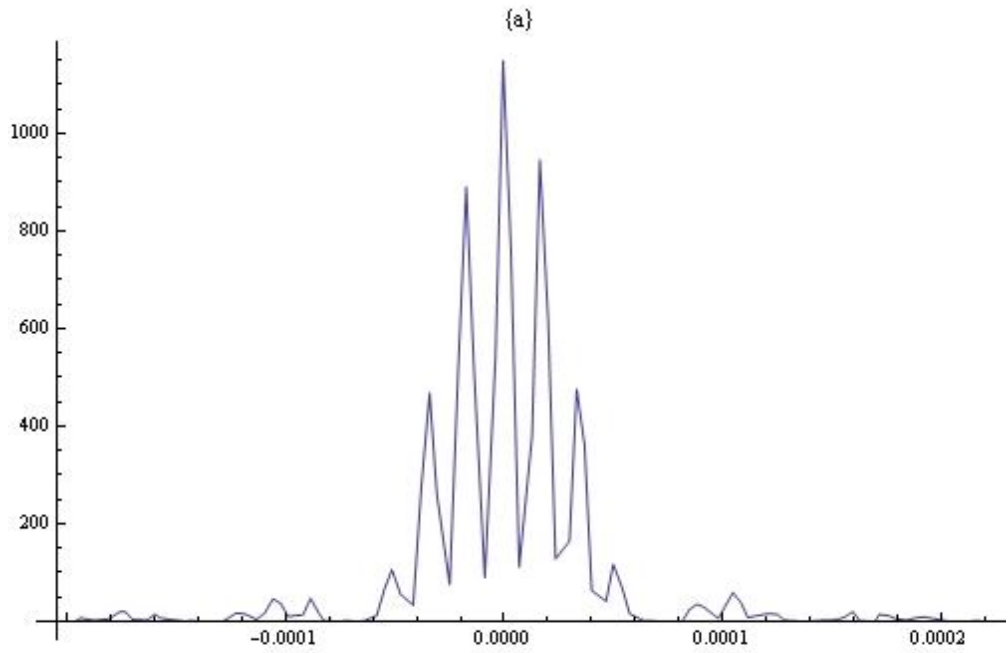

Figure 1(a)

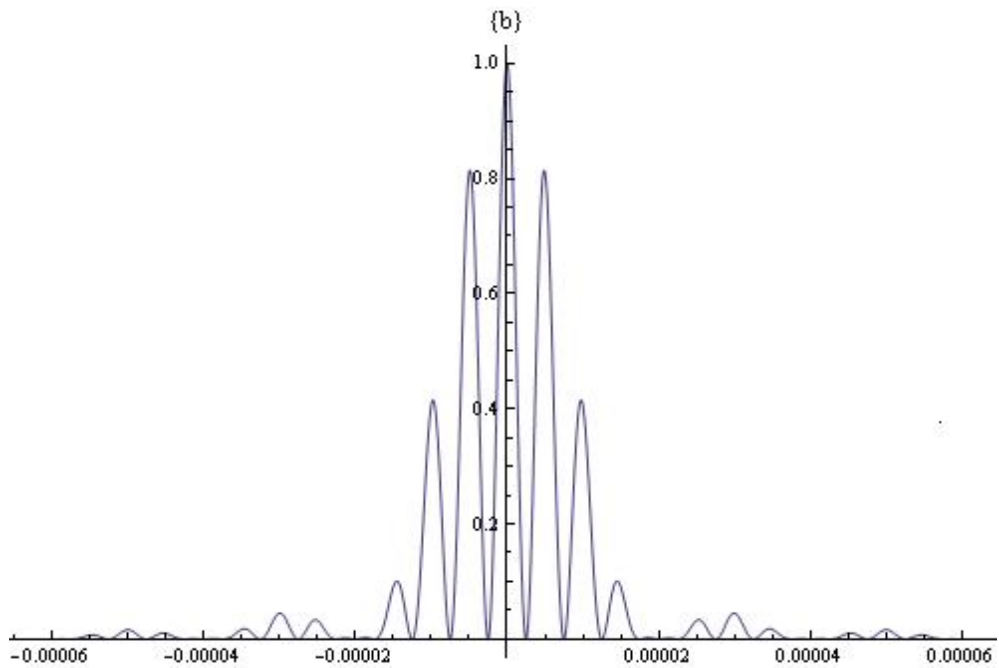

Figure 1(b)

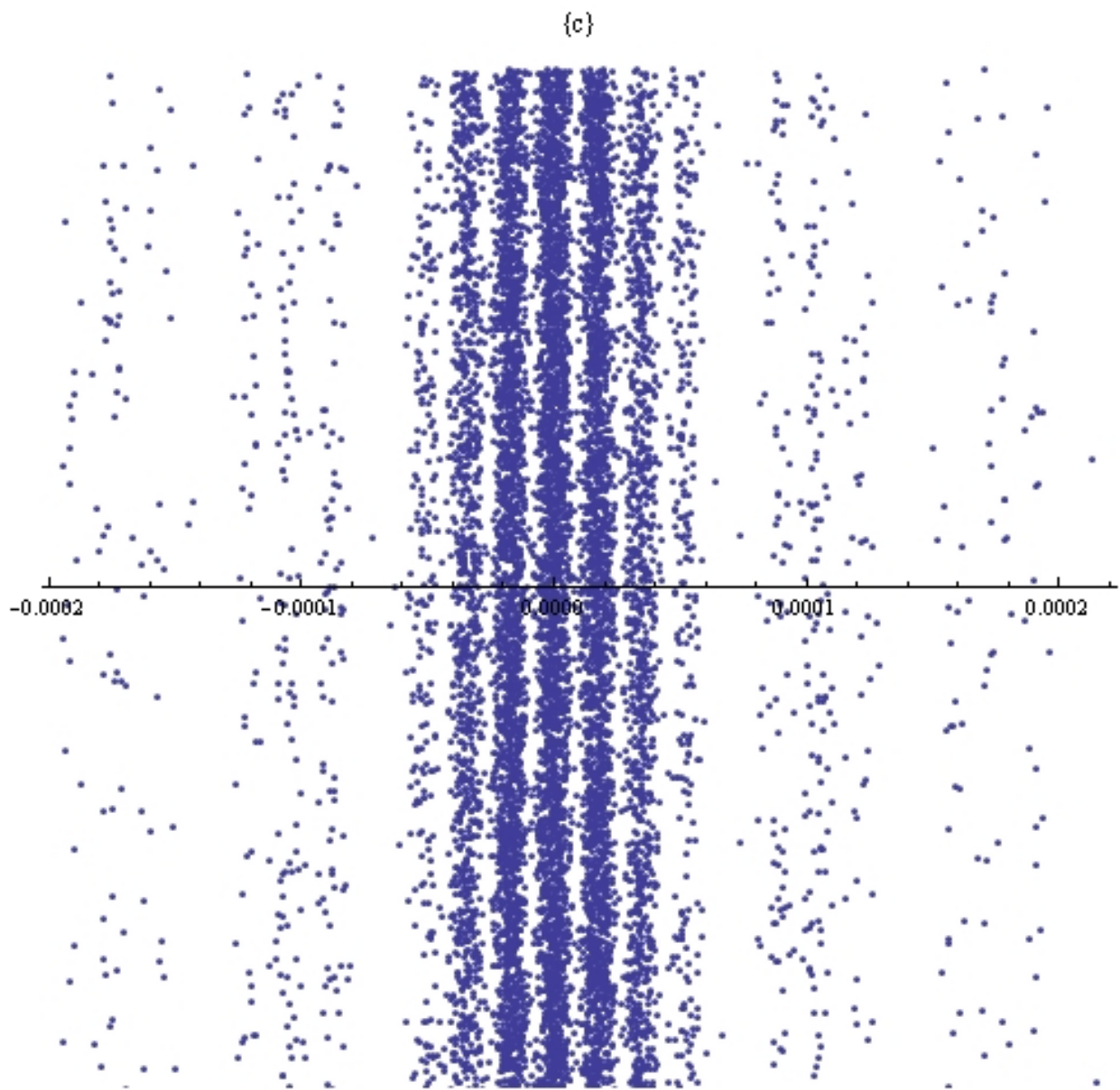

Figure 1(c)

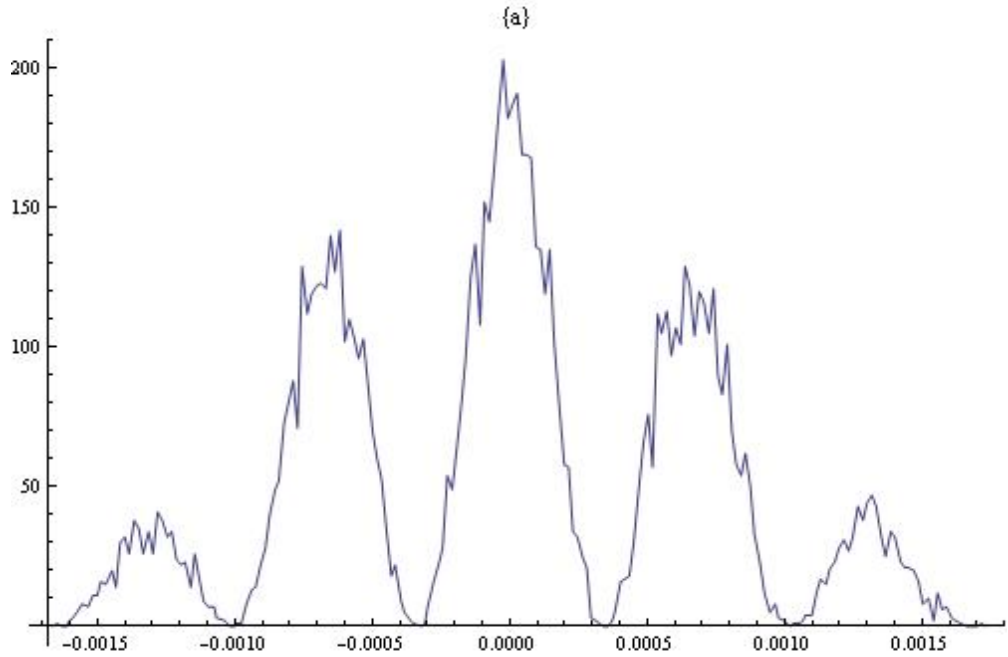

Figure 2(a)

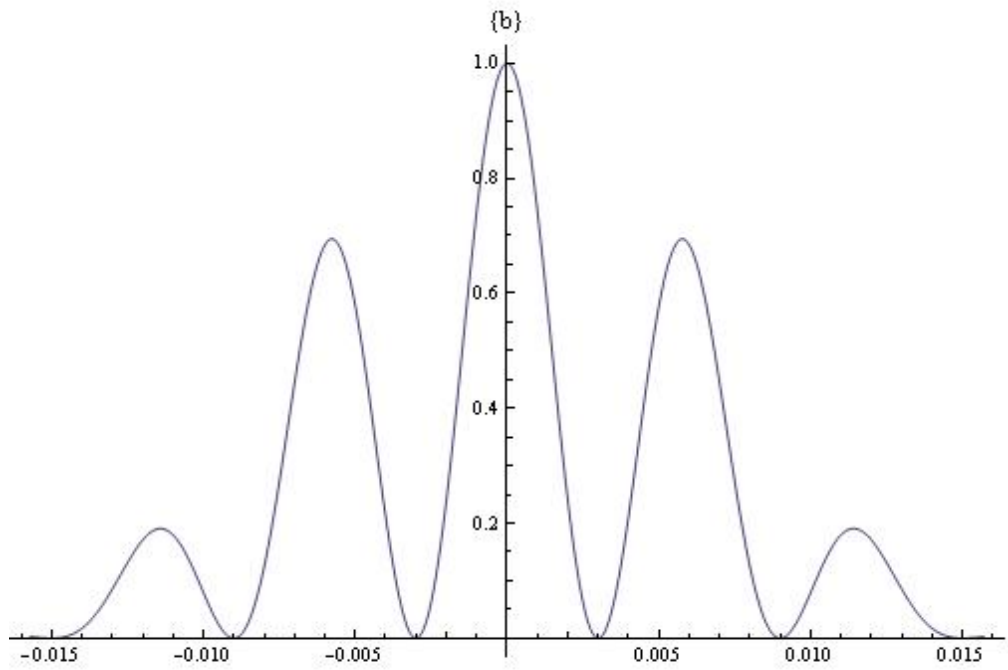

Figure 2(b)

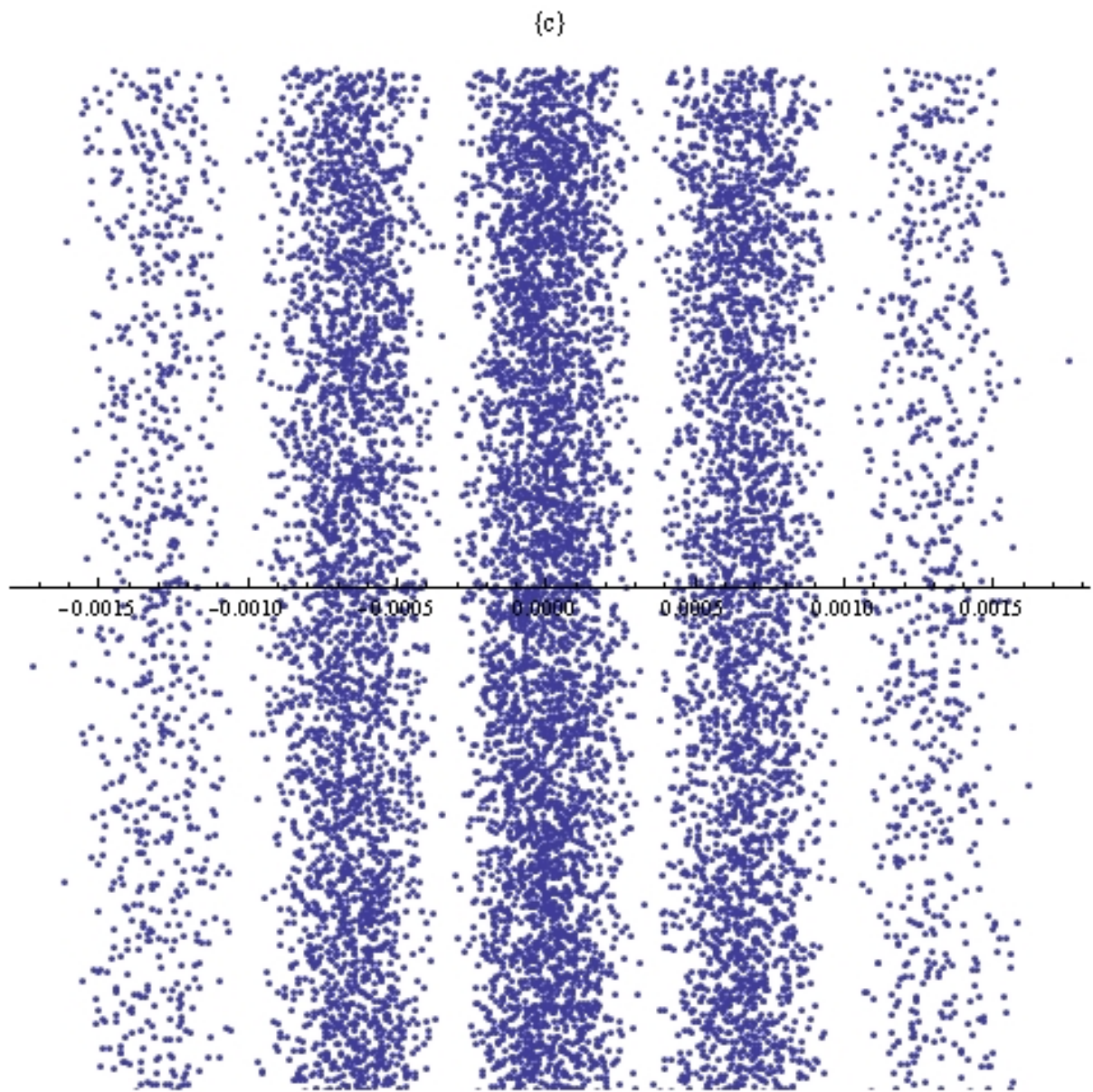

Figure 2(c)